\documentclass[twocolumn,prl,10pt,amsmath,amssymb,nofootinbib,showpacs,superscriptaddress,floatfix]{revtex4-1}
\usepackage{amsmath, mathrsfs, latexsym, amssymb, color, graphicx, comment, hyperref, scrhack}
\usepackage{lmodern}
\usepackage{physics}

\newcommand{\dfr}[2]{\frac {\displaystyle #1}{\displaystyle #2}}

\usepackage{graphicx}
\usepackage{color}
\usepackage[usenames,dvipsnames]{xcolor}

\begin{document}
\title{Glass transition as a topological phase transition}
\author{M.\,G.\,Vasin}
\affiliation{Vereshchagin Institute of High Pressure Physics, Russian Academy of Sciences, 108840 Moscow, Russia}
%\author{V.\,V.\,Brazhkin}
%\affiliation{Vereshchagin Institute of High Pressure Physics, Russian Academy of Sciences, 108840 Moscow, Russia}

\begin{abstract}
The glass transition is considered as a phase transition in the system of topologically protected excitations in matter structure. The critical behavior of the system is considered both in statics and dynamics cases. It is shown in the simple model describing the topological defects system in the elastic medium with non-zero shear modulus, most of characteristic thermodynamic and kinetic properties of glass transition are reproduced: the Vogel--Fulcher--Tammann law; the behavior of susceptibility, and non-linear susceptibilities; heat capacity behavior; and boson peak near the glass transition temperature.
\end{abstract}

%\begin{keyword}
%Glass transition, topological phase transition, Vogel--Fulcher--Tammann law, non-linear susceptibility, Boson peak
%\end{keyword}

\flushbottom
\maketitle
\thispagestyle{empty}

\section{Introduction}

Currently, glasses are widely used in everyday life and technology, and are also the subject of active research in many areas of condensed matter physics. However, the microscopic mechanisms that give rise to this state of matter are still the subject of discussion.

Today there are many various approaches to solving the problem of describing a glass transition. A wide range of ideas is distributed between two extreme points of view.
On the one hand, it is assumed that the glass transition is a purely kinetic effect, result of dynamical locking of liquid disordering structure in the process of rapid quenching~\cite{I1}.
According to this approach glass transition is a pronounced relaxation process obeying kinetic laws. When approaching the liquid-to-glass transition, molecular rearrangements in glass-forming melts become so slow that the change in structure does not have time to follow the decrease in temperature.
The ratio of the structural relaxation time to the melt cooling rate plays decisive role in this case~\cite{UFN1}, and the characteristic feature of the relaxation kinetics of glassforming melts is the dynamic heterogeneity of the structure \cite{DH,DH1}.
The significant progress of the kinetic theories is associated with the approaches based on the mode-coupling theory \cite{Gotze, Sokolov}, in which dynamics is determined by static equilibrium averages. They starts with Newton's equations of motion and ends up with certain experimental predictions \cite{Das}. The  mode-coupling theory predicts a critical temperature below which there is no ergodic phase. The theory is believed to correctly explain the onset of viscous behavior upon cooling, but is not so good at the low temperatures \cite{Dreyfus}.
The opposite hypothesis is that the viscous slowing down is a consequence of an underlying or narrowly avoided phase transition \cite{Sethna,Kivelson}, and the glass phase the result of a genuine thermodynamic phase transition to a disordered solid state \cite{I2,BBL2021}. These theoretical approaches includes one of most popular one today is the random first-order transition theory \cite{Kirkpatrick,Xia}.

This article will provide arguments rather in favor of the second statement.  Although at the liquid-glass transition the system has not any observable order parameter that changes at this transition, nevertheless one can claim that it is one of the form of thermodynamic phase transition. In particular, in this paper it will be shown that the transition to the glass phase can be described as a topological phase transition.

As is known, a topological phase transition is a phase transition between phases whose properties are not explained by standard arguments without involving the topological properties of systems. In our case, it will be considered the phase transition in a system of interacting topological defects (vortices), which leads to the appearance at a finite temperature of a quasi-long order corresponding to a vortex system with an infinite correlation radius. That is, it is a transition between two states of a system of topological defects: a state in which these defects are mobile, and a state in which they are frozen.

The basic idea of this approach is not new. It is founded on the known fact that any liquid is solidlike when probed on a sufficiently short time scale, its instantaneous local structure is similar to the order in solid, and its short-time elastic properties are characterized by the instantaneous elastic moduli. That is the ground of the wide class of elastic models involve assumptions and reasoning which have a long history also in the field of point defects in crystals~\cite{Dure}. The main idea one of classes of these models was offered in number works the late seventies onwards last century~\cite{I6,I7,Fradkin,NR}, and was that physical properties of glasses are governed by a extended constituent, the odd line or disclination which is the only structural element surviving the absence of generative homogeneity and the triviality of the space group~\cite{I4}. The theoretical description of the glass phase as a frozen system of topologically stable defects was actively developed, for example,  in \cite{I3,I4,I5}. One of the most attractive properties of this approach was its universality, allowing to describe glass transitions in various physical systems in terms of common formalism. At around the time, the idea was suggested that the transition to the glass state is a topological phase transition \cite{I8}.

In this paper a further development of this approach is proposed. It is shown that a system of topological defects in a medium with a nonzero momentum shear modulus experiences a phase transition, the theoretical description of which is reduced to a fairly simple statistical model. This phase transition is analogous in nature with the Berezinskii--Kosterlitz--Thouless transition in two-dimensional systems and is one of the forms of topological phase transitions. The presented below analysis of critical properties of this model, including critical dynamics, shows that it is able describe most of universal characteristic properties of glass transitions.

The article is organized as follows: First, a theoretical model based on the theory of elastic deformations is formulated, which takes into account the presence of topologically protected perturbations corresponding to violation of symmetry concerning an axial rotation.
Then the theoretical analysis of the critical behavior of this model in the vicinity of the phase transition in the subsystem of disclinations is carried out.
The paper considers both static and dynamic cases. In the first case, the thermodynamic properties of the phase transition described by the model are calculated, in the second, the frequency dependence of the dynamic structure factor of the system under consideration is determined.

\section{Model}

We consider a liquid in terms of the approach proposed by Yakov Frenkel~\cite{F}.  This approach is based on postulation of some similarity between crystals and liquids, and allows insight into many important properties of the latter. It is based on the assumption that at moderate temperatures, the particles of liquid behave in a manner similar to ones in a crystal. However, while in crystals they oscillate around their nodes, in liquids, after several periods, the particles change their nodes.

The validity of this approach is due to the fact, that in experiments the elastic properties of liquids well noticeable on small scales in space and time. In particular, at high frequencies in liquids, a finite value of the shear modulus and a solid-like oscillation spectrum are observed \cite{B5,B6,B7,B8,B9,B10,B11,B12}. Since transverse phonons can exist in the liquid only at frequencies exceeding the value of the inverse relaxation time, which decreases with the temperature increasing, then the main criterion distinguishing a quasi-gas (soft) fluid from a  solid-like one (Frenkel liquid) one can count the zero value of the shear modulus in entire possible frequency spectrum. On the phase diagram the region of the crossover from one liquid type to other one called Frenkel line \cite{B1,B2}.
Accordingly, the system can be considered solid one if the shear modulus is different from zero on the entire scale of the measured frequencies.
Thus, at low temperatures according to the Frenkel's approach, we consider a liquid as an elastic media containing both elastic and plastic deformations. The plastic deformation presence provides fluidity, and the elastic deformation defines the system free energy.

The free energy density of a deformed elastic system is written as follows:
\begin{gather}\label{I1}
\mathcal{F}=\dfr{\lambda}{2}u_{ll}^2+\mu \hat{\bf u}^2,
\end{gather}
where $\lambda $ is the bulk modulus, $\mu$ is the instantaneous shear modulus, $\hat{\bf u}=u_{ij}={\mathrm{d} u_j}/{\mathrm{d} x_i}=\nabla_iu_j$ is the distortion tensor (${\bf u}$ is the strain vector). It should be noted, that $\mu $ is the microscopic parameter, which corresponds to macroscopic shear modulus in case then the relaxation time of  system structural  surpasses by far the observation time. Since the shear modulus usually depends on the temperature and increases at the temperature decreasing, one can assume that in some temperature interval near glass transition this dependence constitutes a linear function: $\mu=\varepsilon (T_0-T)$, where $T_0$ is some effective temperature parameter. Then it is the temperature at which the shear elasticity appears in the liquid, and which corresponds to the Frenkel line~\cite{B1,B2} on the phase diagram. The condition of the zero value of the average (measured) statical shear modulus of the system in the liquid state is satisfied at the accounting of the presence of movable plastic distortions in this system.

Thus we consider liquid as elastic media, which is fluid because of presence a lot of mobile plastic deformations corresponding to dislocations and disclinations.  In the static consideration the system is in mechanical equilibrium, and the ${\bf u}$ field is free one. However, the elastic energy of this system is not zero since its non-ordering structure is geometrically frustrated and contains a number of stressful regions caused by local topologically protected distortions. The topologically protected rotation distortion corresponds to disclination (or vortex line), and for simplicity below we will consider only this distortion type. Besides, we note since the disclination is caused by a violation of axial symmetry, then, according to topological laws, they are linear objects, and the field describing the interaction between them is Abelian.

Let the system contains a disclination in the point ${\bf r}_n$. It breaks the simple connectivity of the space and leads to appearing in the distortion tensor an irreducible part corresponding to the rotation at movement around this disclination:
\begin{gather}\label{I2}
\oint \hat {\bf u} \mathrm{d}{\bf l}=\int \nabla\times \hat {\bf u}\,\mathrm{d}^2{\bf r}={\bf \Omega}\delta_{{\bf r=r}_n}^{(2)},
\end{gather}
where the space integration is the integration over dimensionless variable ${\bf r}$: $V^{-1}\int\mathrm{d}V=\int\limits_{|{\bf r}|<1}\mathrm{d}^3{\bf r}$, and ${\bf \Omega}$ is the Frank vector. We gutta bear in mind it is a pseudovector.

If the system contains $N$ disclinations, then the its partition function can be represented in the form of the functional integral:
\begin{gather*}
W= \int \mathcal{D}\hat{\bf u}\exp\left[-\beta\int\mathrm{d}^3{\bf r}\,\mathcal{F}\right]\prod\limits_{n=1}^N\delta\left( \nabla\times\hat{\bf u}_{{\bf r}_n}-{\bf \Omega} J_{{\bf r}_n}\right),
\end{gather*}
where $\beta=1/k_bT$, $\delta (\ldots )$ is the functional delta-function, and $J_{\bf r}=\pm 1$ or $0$.
Using the integral representation of the delta-function one can represent the partition function of the system as follows:
\begin{gather*}
W= \iint \mathcal{D}\hat{\bf u}\mathcal{D}{\bf A}\exp\left[-\beta\int\mathrm{d}^3{\bf r}\,\mathcal{H}\right].
\end{gather*}
where ${\bf A}$ is an ancillary field, which forms the condition (\ref{I2}), and the effective Hamiltonian density of the system has the following form:
\begin{gather}\label{A}
\mathcal{H}=\dfr12{\mu}{\hat{\bf u}}^2+
i\beta^{-1}{\bf A}\cdot\left[\nabla\times\hat{\bf u}-{\bf \Omega} \sum\limits_{n=1}^{N}J\delta_{{\bf r=r}_n}^{(2)}\right],
\end{gather}
where $N$ is the quantity of the disclination elements, and ${\bf r}_n$ ($n=1$, 2,$\ldots $, $N$) are their coordinates.
Note that the first term of the free energy (\ref{I1}) is absent here since the rotation tensor is non-diagonal.

After integration over $\hat{\bf u}$-field the system effective Hamiltonian density takes the following form:
\begin{gather}
\mathcal{H}= \dfr{\beta^{-2}}{2\mu}(\nabla\times {\bf A})^2-i\beta^{-1}{\bf \Omega} {\bf A}\sum\limits_{n=1}^{N}J\delta_{{\bf r=r}_n}^{(2)},
\label{Z}
\end{gather}
that corresponds to the system of vortices that interact by the ${\bf A}$-field.
If one takes only two vortices and carries out integration over ${\bf A}$-field, then one can be convinced that this interaction is the coulomb one.

We consider the case then the vortices number is arbitrary, therefore we can carry out averaging over the grand canonical ensemble of the vortices (see Appendix I). After this averaging the system effective Hamiltonian density assumes the form:
\begin{gather}
\mathcal{H}= \dfr{\beta^{-2}}{2\mu}(\nabla\times {\bf A})^2-g\beta^{-1}\cos\left({\bf \Omega  A}\right),
\label{Z1}
\end{gather}
where  $g$ is the vortices density, and which is nothing but the Hamiltonian density of the sine-Gordon theory\,\cite{I9,Minnhagen}.

Let us expand the cosine term into the power series. The quantum field theory teaches us that in $d$ dimension systems close to the critical point only the terms of the Taylor series expansion with powers of ${\bf A}$ less than $2d/(d-2)=6$ are relevant\,\cite{Zee}. It means that in the 3D case only the first two terms of this expansion are relevant and the third one is marginal. Thus the fluctuation corrections are relevant only for these first two terms, and the system effective Hamiltonian (\ref{Z1}) density can be represented as follows:
\begin{multline*}
\mathcal{H}=\dfr{\beta^{-2}}{2\mu}(\nabla\times {\bf A})^2\\
+g\beta^{-1}\left({\bf \Omega A} \right)^2\left(\dfr{1}{2}-\dfr{\left({\bf \Omega A}\right)^2}{4!}+\dfr{\left({\bf \Omega A}\right)^4}{6!}\right).
\end{multline*}

Let us separate the ${\bf A}$ field on fast, $\tilde{\bf A}$, and the slow, ${\bf A}$, parts: ${\bf A}\to {\bf A}+\tilde{\bf A}$, and average the model over $\tilde{\bf A}$.
Taking into account that $\langle \tilde{\bf A}\rangle=0$ we can rewrite the above expression as follows:
\begin{multline}\label{O1}
\mathcal{H}=\dfr{\beta^{-2}}{2\mu}(\nabla\times {\bf A})^2+g\beta^{-1}\dfr{\Omega^2}{2}{\bf A}^2\left(1-\dfr{\Omega^2 }{2} \langle \tilde{\bf A}\tilde{\bf A}\rangle_0\right)\\
-g\beta^{-1}\dfr{\Omega^4}{4!}{\bf A}^4\left(1-\dfr{\Omega^2 }{2} \langle \tilde{\bf A}\tilde{\bf A}\rangle_0\right)+g\beta^{-1}\dfr{\Omega^6}{6!}{\bf A}^6,
\end{multline}
where
\begin{gather}\label{O2}
\langle \tilde{\bf A}\tilde{\bf A}\rangle_0=\lambda_D\int\limits_0^{\lambda_D^{-3}}\dfr{\mathrm{d}^3{\bf q}}{(2\pi)^3}\dfr{\mu\beta}{{\bf q}^2}=\int\limits_0^{1}\dfr{\mathrm{d}^3{\bf p}}{(2\pi)^3}\dfr{\mu\beta}{{\bf p}^2}= \dfr{\mu\beta}{2\pi^2},
\end{gather}
$\lambda_D$ is the Debye length, and ${\bf p}$ is the dimensionless momentum, ${\bf p}=\lambda_D{\bf q}$.
It leads to the following effective Hamiltonian density presentation:
\begin{multline}
\mathcal{H}=\dfr{\beta^{-2}}{2\mu}(\nabla\times {\bf A})^2\\
+M^2\left(\dfr12(\Omega{\bf A})^2
- \dfr1{4!}(\Omega{\bf A})^4 \right) +g\beta^{-1}\dfr{\Omega^6}{6!}{\bf A}^6,
\label{EfH}
\end{multline}
where $M^2 =g\Omega^4\varepsilon (T-T^*)/(2\pi)^2$ is the square of the effective ${\bf A}$ field ``mass'', and
\begin{gather}\label{Tg}
T^*=\dfr{T_0}{1+\dfr{k_b}{\varepsilon}\left(\dfr{2\pi}{\Omega}\right)^2 }
\end{gather}
is the phase transition temperature in the disclination subsystem. It is important that from the above derivation the $T^*$ is proportional to the shear modulus of the glass state in $T^*$, $T^*= \mu(T^*) (\Omega/2\pi)^2/k_b$, that is known experimental fact~\cite{Wang}. It is means that $T^*$ is the well determined thermodynamic value of the condensed system, which directly relates with its elastic properties.

In order to understand what is the gist of this phase transition one need to revisit the (4) expression from which one can see that the ${\bf A}$-field is the interaction field between the $J$-particles.  One can see this field is analogous to the vector potential of current's electromagnetic field with the only difference that it is massive. As a result the interaction between vortices is short-range, and their system is the fluid system of loosely-coupled particles. The phase transition occurs at $T=T^*$, when the ${\bf A}$-field becomes massless, and corresponding interaction becomes long-range. The net-like phase of strong-coupled disclinations forms.
One can see that this is a topological phase transition. By definition topological phase transition is a phase transition between phases whose properties are not explained by standard arguments without involving the topological properties of systems. In our case, this is a phase transition in a system of interacting topological defects (vortices), which leads to the appearance at a finite temperature of a quasi-long order corresponding to a vortex system with an infinite correlation radius. I.e., it is a transition between two states of a system of topological defects: a state in which these defects are mobile, and a state in which they are frozen.

The ${\bf A}$ field corresponds to the gauge field of the vortex--vortex interaction in \cite{NR,I4,I3,I5,I8,I9}. This field is arbitrary, therefore one can choose it in Coulomb gauge, $\nabla \cdot{\bf A}=0$.
Then the correlation function of the ${\bf A}$ field corresponding to the above Hamiltonian is
\begin{gather*}
\langle {\bf AA}\rangle_{\bf p}=\dfr{\beta }{\mu^{-1}{\bf p}^2+\beta^2 M^2 }.
\end{gather*}
It is not difficult to check that the static shear modulus $\mu_{st}\sim \langle {\bf AA}\rangle_{{\bf r}=0}=0$ when $T>T^*$, and  $\mu_{st}= \mu(T^*)\neq 0$ in $T=T^*$. Thus, from the definition of difference between liquid and glass states it means that $T^*$ can be interpreted as the liquid-to-glass transition temperature.

One should recalls that the above description of the system is correct only in the temperature region $T_0< T\geqslant T^*$, where the disclination system can be described in framework of equilibrium statistical mechanics.

\section{Critical behavior of the vortex system near $T^*$}

In the pulse space the pair correlation function of the vortices is
\begin{gather*}
\langle {\bf J}{\bf J}\rangle_{\bf p} \propto\exp\left[\dfr{\Omega^2\beta}{\mu^{-1}{\bf p}^2+\beta^2 M^2 }\right]
\end{gather*}
(see Appendix II).
This function allows estimate correlation length of the vortices with opposite charges. Indeed, if the vortex correlation $\langle {\bf J}{\bf J}\rangle$ has a correlation length $L_c$, i.e. it can be presented as $|\langle {\bf J}{\bf J}\rangle|_{\bf r}\propto \exp (-|{\bf r}|/L_c)$, then
\begin{gather*}
L_c\propto |\langle {\bf J}{\bf J}\rangle|_{p=0}^{1/3}=\exp\left[\dfr{k_b}{3g\varepsilon}\left(\dfr{2\pi}{\Omega}\right)^2\dfr{T}{T-T^*}\right]
\end{gather*}
(see Appendix III).

At first glance, it may seem strange that the vortex correlation length increases according to the faster law than the ${\bf A}$-field correlation length. However, it is not difficult to understand this, if remember that the ${\bf A}$-field describes the interaction between vortices. Therefore, the correlation length of the ${\bf A}$-field is the radius of interaction between vortices, which is not the correlation length of the vortices but is associated with it by more complex nonlinear relations. An increase of this radius leads to an increase in the effective number of nearest neighbors of each vortex, which reduces the percolation threshold and dramatically increases the size of the percolation region which is much greater than this radius. Thus the vortices correlation length grows faster than the interaction radius.

Unfortunately, vortex correlation length, as well as $\langle {\bf J}{\bf J}\rangle$ and $\langle{\bf AA}\rangle$ correlation functions cannot be directly observed in an experiment. However, there are other quantities, connected with them, which can be directly measured. First of all, from the critical dynamics, it follows that the vortex correlation length connected with the system's relaxation time, $\tau\propto L_c^z$, where $z$ is the dynamical exponent. Therefore
\begin{gather}\label{VFT}
\tau\propto\exp\left[\dfr{zk_b}{3g\varepsilon}\left(\dfr{2\pi}{\Omega}\right)^2\dfr{T}{T-T^*}\right],
\end{gather}
that is nothing but the Vogel--Fulcher--Tammann (VFT) law for the temperature dependence of the relaxation time near the glass transition point, and $T^*$ corresponds to the Vogel--Fulcher temperature being determined in experiment.
At first sight, this seems to be surprising since the VFT law is characteristic of frustrated systems near the glass transition. However, this is quite well natural since the vortex--vortex interaction, described by the ${\bf A}$-field, like the coulomb one is long-range, and the systems with a long-range interaction are frustrated \cite{LR-Fr1,LR-Fr2}. Thus, the description of the interaction in terms of the gauge field ${\bf A}$ allows using the methods of equilibrium statistical mechanics for the description of systems with long-range interaction.

\subsection{Susceptibility}

\begin{figure}[h!]
   \centering
   \includegraphics[scale=1]{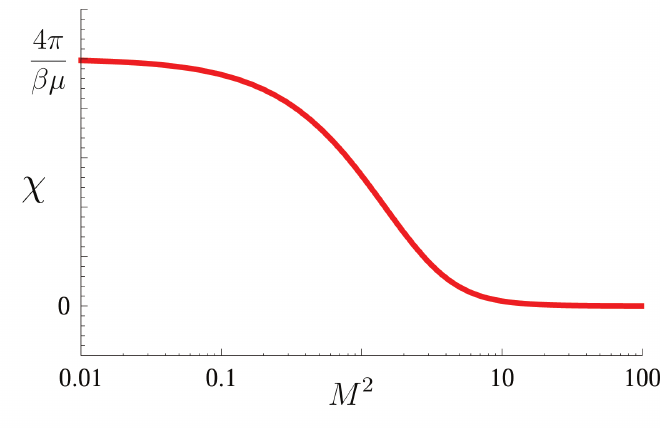}
   \caption{Graphical illustration of the susceptibility change at the glass transition in the presented theoretical description (with a logarithmic $M^2$ axis).}
  \label{Sus}
\end{figure}

In order to derive the expression for linear and nonlinear susceptibilities of the system near $T^*$, let us follow the standard way and add to our system an external force, thus the system's Hamiltonian is rewritten as follows:
\begin{gather}\label{S1}
\mathcal{H}=\dfr{\mu}2{\hat{\bf u}}^2+
i\beta^{-1}{\bf A}\cdot\left[\nabla\times\hat{\bf u}-{\bf \Omega} \sum\limits_{n=1}^{N}J\delta_{{\bf r=r}_n}^{(2)}\right]+\hat{\bf \sigma}\cdot\hat{\bf u},
\end{gather}
where $\hat{\bf \sigma}$ is the externally induced stress tensor per unit volume, which after the calculation is supposed to be sought zero.
The linear susceptibility can be derived by the differentiation: $\chi=\partial_{\bf f}\langle{\bf u}\rangle$, where ${\bf f}=\nabla \cdot\hat\sigma$ is an external force. One can ascertain it is proportional to the $\langle {\bf u}{\bf u}\rangle_{{\bf p}=0}=\left.{\bf p}^{-2}\langle \hat{\bf u}\hat{\bf u}\rangle_{\bf p}\right|_{{\bf p}=0}$ correlation function. Taking into account that $\hat{\bf u}$ is the antisymmetric tensor corresponding to the local rotation of the media on some angle $\omega$, $\omega_i=\varepsilon_{ijk}u_{jk}$, and using the manipulations presented in Appendix IV, one finds that:
\begin{multline*}
\langle {\bf u}{\bf u}\rangle_{\bf p}
= \dfr1{{\bf p}^2}\left[ \left(\dfr{1}{\beta\mu}\right)^2{\bf p}^2\langle {\bf AA}\rangle_{\bf p}+\dfr{1}{\beta\mu}\right]\\
=\dfr{1}{\beta\mu}\left(\dfr {\mu^{-1}}{\mu^{-1}{\bf p}^2+\beta^2 M ^2}+\dfr1{{\bf p}^2}\right).
\end{multline*}
As a result one can write the expression for susceptibility as follows:
\begin{multline*}
\chi=\int\limits_{0}^{1}\mathrm{d}^3{\bf r}\langle {\bf u}{\bf u}\rangle_{\bf r}
=\dfr{4\pi}{\beta\mu}\int\limits_{0}^{1}\mathrm{d}r\,r\left(\exp(-r\sqrt{\mu}\beta M ) +1\right)\\
=\dfr{4\pi}{\beta\mu}\left(\dfr12+\dfr{1-(1+\sqrt{\mu}\beta M )\exp(-\sqrt{\mu}\beta M )}{\mu \beta^2M^2}\right).
\end{multline*}
One can see that near the $T^*$, when $M \to 0$, this value does not divergent (see Fig.\,\ref{Sus}). This is in agreement with experimental observations and points that the glass transition is not a second-order phase transition.

\subsection{Non-linear susceptibilities}

In the glasses, the particularly interesting are third- and fifth-order susceptibilities. It is supposed that their divergence at the glass transition temperature is evidence that the glass transition is a phase transition, and points to the growth of thermodynamic amorphous order in glass-formers.

The third-order non-linear susceptibility $\chi_3$ is proportional to the quadruple correlator of the ${\bf u}$-field: $\chi_3=\partial^3_{\bf f}\langle{\bf u}\rangle_{{\bf p}=0}\propto\langle{\bf u}^4\rangle_{{\bf p}=0}$ which can be represented in the following form (see Appendix IV):
\begin{figure}[h!]
   \centering
   \includegraphics[scale=1]{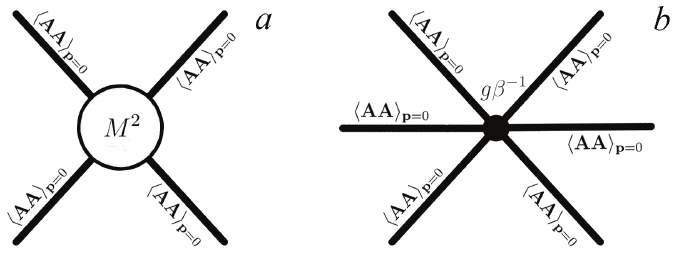}
   \caption{a) The four bold lines in the graph correspond to the correlation functions $\langle {\bf A A}\rangle_{{\bf p}}$ which cross in the vertex, $M^2$, corresponding to the ${\bf A}^4$-term of the effective Hamiltonian. b) Six corresponding to the correlation functions graphs cross in the vertex, $g\beta^{-1}$, corresponding to the ${\bf A}^6$-term.}
  \label{cor4}
\end{figure}
\begin{gather*}
\langle {\bf u}^4\rangle_{{\bf p}=0}=\left.{\bf p}^{-4}\langle\hat{\bf u}\rangle_{{\bf p}}\right|_{{\bf p}=0}= \left(\dfr{1}{\beta\mu}\right)^4\langle {\bf A}^4\rangle_{{\bf p}=0}+3\langle{\bf uu}\rangle^2_{{\bf p}=0}.
\end{gather*}
Using the above susceptibility calculation, one can see that second term in this expression gives finite contributions in contrast to the first term which diverges at $T\to {T^*}^+$. Indeed,
the irreducible part of the quadruple correlator $\langle {\bf A}^4\rangle$, which graphically is represented in Fig.\,\ref{cor4}, is written as follows:
\begin{gather*}
\langle {\bf A}^4\rangle_{{\bf p}=0}\propto\left(\dfr{1}{M ^2}\right)^4M ^2.
\end{gather*}
As a result $\chi_{3}\propto\langle {\bf A}^4\rangle_{{\bf p}=0}\propto (T-T^*)^{-3}$.
Similarly, the fifth-order susceptibility is presented as follows:
\begin{gather*}
\chi_{5}\propto \langle {\bf A}^6\rangle_{{\bf p}=0}\propto\left(\dfr{1}{M ^2}\right)^6\propto \dfr1{(T-T^*)^6}.
\end{gather*}
This result is consistent with $\chi_5\propto \chi_3^2$ and supports a picture of amorphous compact domains mostly independent of differences at the molecular level~\cite{I2,BBL2021}.

\subsection{Heat capacity}

\begin{figure}[h!]
   \centering
   \includegraphics[scale=1]{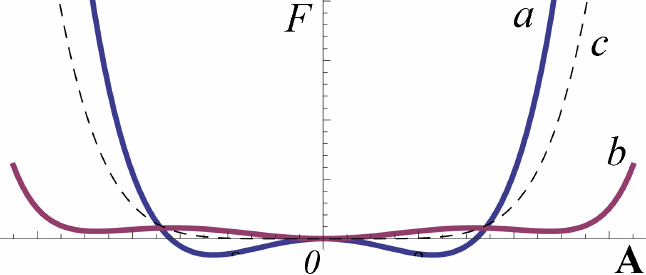}
   \caption{Thermodynamic potential of the system with Hamiltonian (\ref{EfH}) at the temperatures near $T^*$  a) at the temperatures $T<T^*$, b) at the temperatures $T>T^*$, c) at $T=T^*$.}
  \label{Pot}
\end{figure}

First of all, we note that phase transition in the ${\bf A}$-field subsystem is the weak first-order phase transition. The form of the corresponding thermodynamic potential (see Fig.\,\ref{Pot}) evidences the differences in the critical behavior of the system above and below the glass transition temperature $T^*$. From Fig.\,\ref{Pot} one can see that at the approaching to the $T^*$ from above the systems criticality corresponds to a first-order phase transition. While as at the approach to the $T^*$ from below the transition demonstrates the characteristic properties of the second-order phase transition. Thus, one can expect that at $T\to {T^*}^-$ the heat capacity temperature dependence of the considered system will demonstrate sharp pick as well as at the continuous second-order phase transition, but at $T\to {T^*}^+$ it will be a finite jump. This system's behavior qualitatively describes experimental observations of the heat capacity near a glass transition.

Knowing the partition function $W$ one can estimate the vortex subsystem contribution to the heat capacity. At the approach to the $T^*$ from below it is written as follows (see Appendix V):
\begin{gather*}
\Delta C_p=\dfr1{\beta}\dfr{\partial^2\left( T\ln W\right)}{\partial T^2}\propto \dfr1{|T^*-T|^{1/2}}.
\end{gather*}
One can see it diverges as well as when at a second-order phase transition.

\begin{figure}[h!]
   \centering
   \includegraphics[scale=0.8]{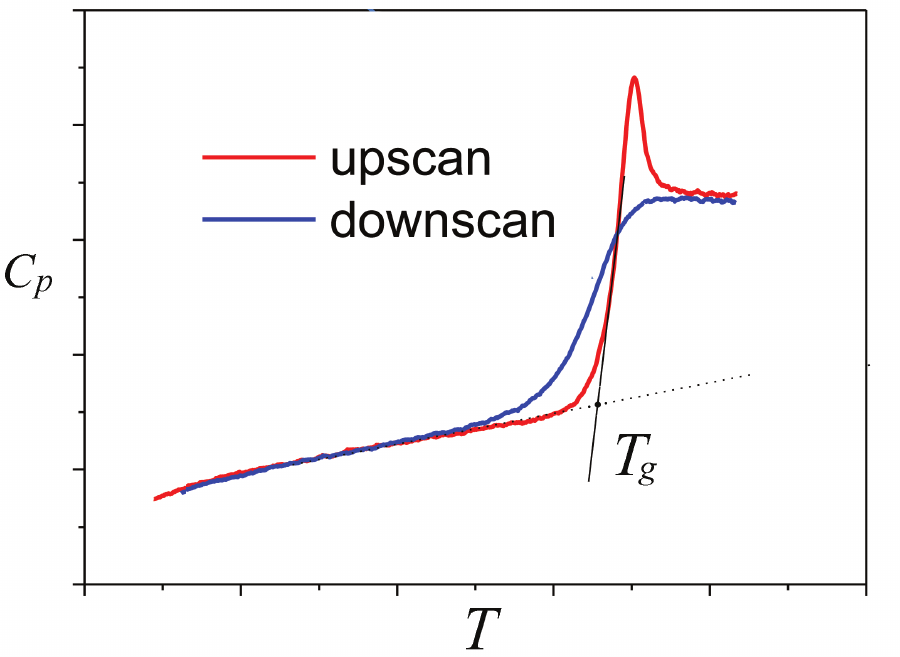}
   \caption{The qualitative picture of the heat capacity behaviour near the glass transition. The blue line corresponds to cooling, and red line to the heating.}
  \label{HC}
\end{figure}

At the approach to the $T^*$ from above the heat capacity behavior is different from the above. This case corresponds to the first-order phase transition in the ${\bf A}$-field subsystem,  which can happen in the temperature interval from $T=T_0$, when the energy of symmetrical and antisymmetrical states of the ${\bf A}$-field subsystem are equal (binodal), till $T\approx T^*$  when the energy barrier between these states vanishes (spinodal) (see Fig.\,\ref{Pot}).
In this case the heat capacity has not singularity but undergoes a jump down at $T^*<T<T_0$. This behavior agrees with experimental observations, which can be illustrated by the picture presented in Fig.\,\ref{HC}. The pulling of the liquid-glass transition temperature to the lower temperatures region, which presences in the experiments and in the figure, can be explained by the well-known phenomenon of the dependence of the temperature of the liquid-glass transition on the quenching rate. One can show this is the result of the anomaly slowing of the relaxation processes near $T^*$ (see for example \cite{V}).
A detail study this problem is possible within dynamic theory which will be considered below.

\section{Critical dynamics and Boson peak}

Usually, the boson peak is observed in the dynamic structure factor of supercooled liquids at temperatures below $T\approx 1.2\,T_g$ ($T_g$ is the glass transition temperature). Its appearance is related to the collective (dynamically correlated) motion of atoms and the formation of clusters with locally favored structures in the normal (quasi-gas) liquid medium.
In the considered our model the vortex formation own implies the presence of a locally ordered structure in which the vortex arises. Therefore, the size of the correlated clusters of this favored structure is defined by the vortices correlation length and ${\bf A}$-field interaction radius. Thus, one can suppose that features of the low-frequency part of the absorption spectrum are related to the ${\bf A}$-field mass renormalization.

In order to describe the boson peak phenomenon in terms of discussed above approach let us use a non-equilibrium dynamics technique. However, before we turn to the Keldysh--Schwinger technique one should keep in mind that that above during averaging over $\hat{\bf u}$ field the system was supposed thermalized. Therefore, the subsequent consideration is correct in supposing that the ${\bf A}$ field is much slower than the $\hat{\bf u}$ field. That seems to be natural.

\begin{figure}[h!]
   \centering
   \includegraphics[scale=0.6]{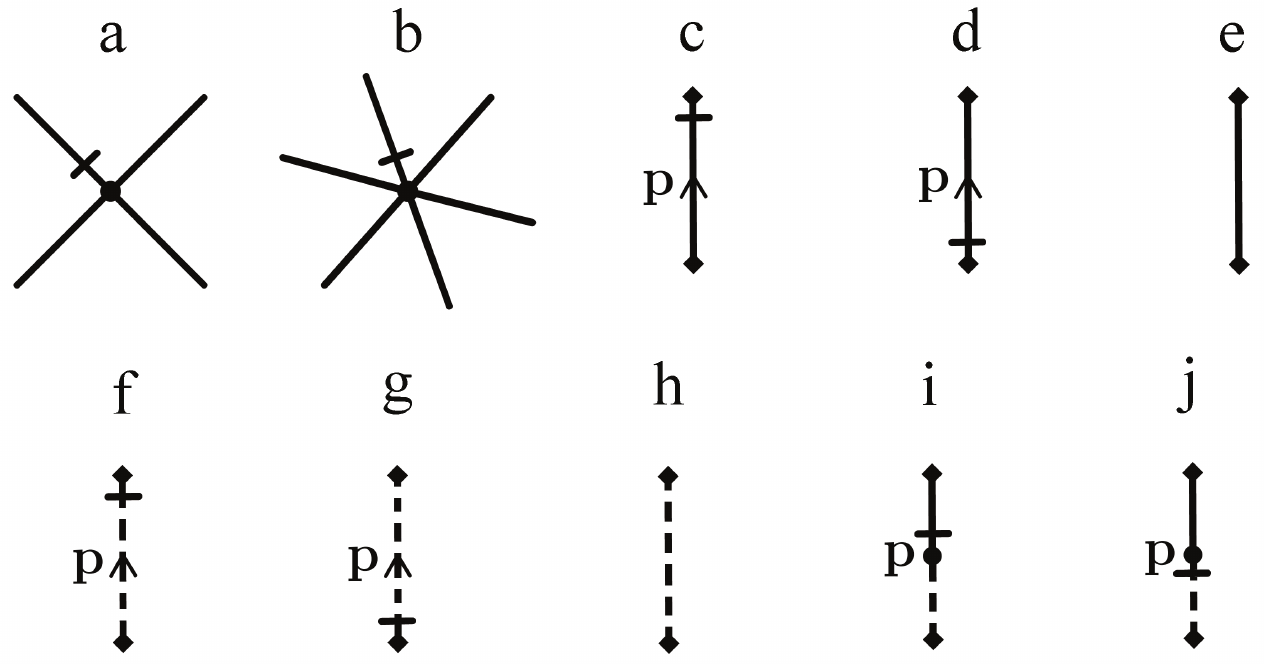}
   \caption{The graphical presentation of the vertices ${\bf A}^3{\bf A'}$ (a) and ${\bf A}^5{\bf A'}$ (b), $[{\bf p}\times\hat{\bf u}']{\bf A}$ (j), $[{\bf p}\times\hat{\bf u}]{\bf A'}$ (i), and the prime Green functions $G^A_0$ (c), $G^R_0$ (d), $G^K_0$ (e), $\Delta^A_0$ (f), $\Delta^R_0$ (g), and $\Delta^K_0$ (h).}
  \label{Prop}
\end{figure}

Same way as above let us separate the $\hat{\bf u}$ and ${\bf A}$ fields on fast, $\tilde{\bf u}$, $\tilde{\bf A}$, and the slow, $\hat{\bf u}$, ${\bf A}$, parts: $\hat{\bf u}\to \hat{\bf u}+\tilde{\bf u}$, ${\bf A}\to {\bf A}+\tilde{\bf A}$. Let us suppose the fast parts of the fields are in thermal equilibrium, and average the model over $\tilde{\bf u}$ and $\tilde{\bf A}$ like in (\ref{EfH}).
Then, taking into account that $\langle \tilde{\bf A}\rangle=0$, we can present the Hamiltonian of our system as follows:
\begin{multline}
\mathcal{H}=\dfr12{\mu}{\hat{\bf u}}^2+
i{\bf A}\cdot\left[\nabla\times\hat{\bf u}\right]+\dfr{1}{2\mu}(\nabla\times {\bf A})^2+\\
M^2\left(\dfr12(\Omega{\bf A})^2
- \dfr1{4!}(\Omega{\bf A})^4 \right) +g\dfr{\Omega^6}{6!}{\bf A}^6.
\label{EfH1}
\end{multline}
If we go over to the dynamic, in terms of the Keldysh--Schwinger technique \cite{Schwinger,Keldysh,KAMENEV,Milton} one can write the system's action functional in the following form:
\begin{multline}\label{L}
\mathcal{S}= \dfr12\bar{\bf u}\hat{\Delta}_0^{-1}\bar{\bf u}+ \dfr{1}{2}\bar{\bf A}{\hat{\bf G}_0}^{-1}\bar{\bf A} +
i{\bf A}\cdot\left[\nabla\times\hat{\bf u'}\right]+i{\bf A'}\cdot\left[\nabla\times\hat{\bf u}\right]\\
- \gamma^{-1}M^2\dfr{2\Omega^4}{4!}{\bf A}^3{\bf A}'  +g\dfr{2\Omega^6}{6!}{\bf A}^5{\bf A}',
\end{multline}
where ${\bf u}$ and ${\bf u}'$ are the ``quantum'' and ``classical'' fields  after Keldysh rotation \cite{Keldysh,KAMENEV}, $\bar{\bf u}$ is the vector  $\bar{\bf u}=(\hat{\bf u},\,\hat{\bf u}')$, ${\hat{\Delta}}_0$ is the prime Green function matrix which in pulse presentation has the following form:
\begin{equation}\label{DF}
\begin{array}{l}
\hat\Delta_0=\left[ \begin{array}{cc} \displaystyle {\Delta^K_0}_{{\bf p},\,\omega} & {\Delta_0^R}_{{\bf p},\,\omega} \\[14pt]
\displaystyle {\Delta^A_0}_{{\bf p},\,\omega} &  0 \end{array}\right]=\left[ \begin{array}{cc} \displaystyle \dfr{2\gamma k_bT}{\mu^2+\gamma^2\omega^2} & \displaystyle \dfr{\gamma}{\mu-i\gamma\omega} \\[12pt]
\displaystyle \dfr{\gamma}{\mu+i\gamma\omega} &  0 \end{array}\right],
\end{array}
\end{equation}
where $\gamma $ is the kinetic coefficient,  $\Delta^R_0,\,\Delta^A_0,\,\Delta^K_0$ are respectively its retarded, advanced, and Keldysh components of $\hat{\Delta}_0$. Similarly ${\bf A}$ and ${\bf A}'$ are the ``quantum'' and ``classical'' fields after Keldysh rotation, $\bar{\bf A}$ is the vector $({\bf A},\,{\bf A}')$, and $\hat{\bf G}_0$ is the prime Green function matrix which in pulse presentation has the following form:
\begin{equation}\label{GF}
\begin{array}{l}
\hat{\bf G}_0=\left[ \begin{array}{cc} \displaystyle {G^K_0}_{{\bf p},\,\omega} & {G^R_0}_{{\bf p},\,\omega} \\[14pt]
\displaystyle {G^A_0}_{{\bf p},\,\omega} &  0 \end{array}\right]\\
=\left[ \begin{array}{cc} \displaystyle \dfr{2\gamma k_bT}{(\gamma^2\mu^{-1}{\bf p}^2+M^2)^2+\gamma^2\omega^2} & \displaystyle \dfr{\gamma}{\gamma^2\mu^{-1}{\bf p}^2+M ^2-i\gamma\omega} \\[12pt]
\displaystyle \dfr{\gamma}{\gamma^2\mu^{-1}{\bf p}^2+M^2+i\gamma\omega} &  0 \end{array}\right],
\end{array}
\end{equation}
where $G^R_0,\,G^A_0,\,G^K_0$ are respectively its retarded, advanced, and Keldysh components (see Fig.\,\ref{Prop}).

\begin{figure}[h!]
   \centering
   \includegraphics[scale=0.7]{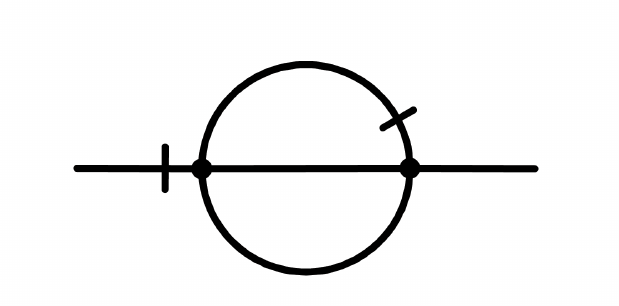}
   \caption{This term gives logarithmically divergent contribution to the $M^2 $ renormalization.}
  \label{Sigma}
\end{figure}
Near to the glass transition the ${\bf A}$-field full correlation function matrix, $\hat G_{{\bf p},\,\omega}$, differs from the prime one by the addition of the self-energy term, $\Sigma_{{\bf p},\,\omega}$, to the ${\bf A}$-field mass: $M^2\,\rightarrow \,{M_R^2}_{{\bf p},\,\omega}=M^2-\gamma\Sigma_{{\bf p},\,\omega}$.
Using the perturbation theory one can find the dominant near $T^*$ contribution to the self-energy is given by the logarithmically divergent term
\begin{multline*}
  \Sigma_{{\bf p},\,\omega}\approx\dfr{M^4 \Omega^4}{8\gamma^2}\iint \dfr{\mathrm{d}^3{\bf p'}\mathrm{d}\omega'}{(2\pi)^4}\dfr{\mathrm{d}^3{\bf p''}\mathrm{d}\omega''}{(2\pi)^4}\times\\
   G^K_{{\bf p'},\,\omega'}G^K_{{\bf p''},\,\omega''}G^{A(R)}_{{\bf p'}+{\bf p''}+{\bf p},\,\omega'+\omega''+\omega},
\end{multline*}
which is graphically presented in Fig.\,\ref{Sigma}. In the assuming of small $M^2 $ in the ${\bf p}\to 0$ limit (see Appendix VI)
\begin{gather*}
\Sigma_{{\bf p}\approx 0,\,\omega}\approx M^4\dfr{\mu^3}{\beta^2\gamma^7}\dfr{\Omega^4}{8(2\pi)^2}\log\omega.
\end{gather*}

\begin{figure}[h!]\label{Prop2}
   \centering
   \includegraphics[scale=0.65]{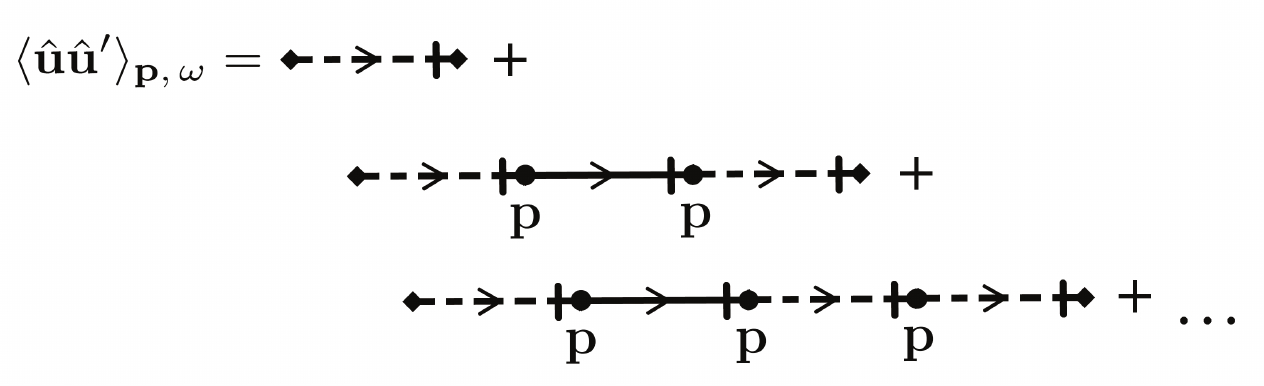}
   \caption{Graphical presentation of the retarded component of the $\hat{\bf u}$-field Green function, which is a geometrical progression.}
\end{figure}
\begin{figure}[h!]
   \centering
   \includegraphics[scale=0.9]{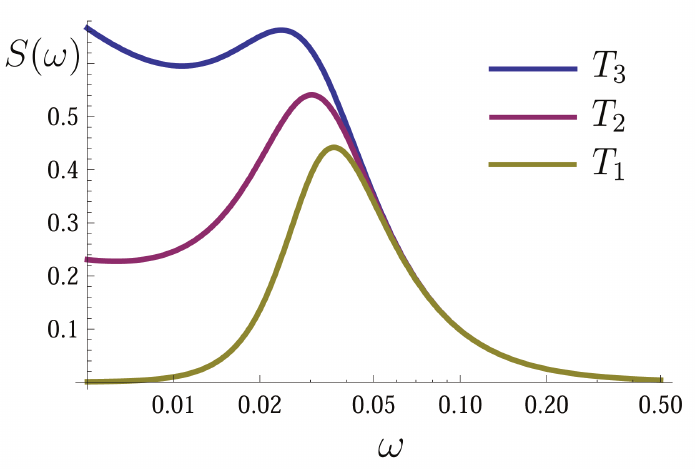}
   \caption{The qualitative form of the frequency dependence of the dynamic structure factor, $S(\omega )$, derived with (\ref{DSF}) expression, with a logarithmic frequency axis ($T_1\approx T_g<T_2<T_3$). With temperature growth the peak maximum position shifts to low frequencies, and its height and width growth.  At $T\to T^*$ the central peak vanishes. This picture qualitatively well agrees with experimental data (see, for example, \cite{Chem}). }
  \label{BP}
\end{figure}

In order to determine the expression for the dynamic structure factor, $S(\omega)$, one should determine the Keldysh part of the time-dependent full correlation function matrix $\Delta^K_{\omega}=\langle \hat{\bf u}\hat{\bf u}\rangle_{\omega} \propto S(\omega)$. In pulse representation the retard part of this matrix is written  as follows (see Fig.\,\ref{Prop2}):
\begin{multline*}
\Delta^R_{{\bf p},\,\omega}={\Delta^A}^*_{{\bf p},\,\omega}=\langle {\hat{\bf u}\hat{\bf u}'}\rangle_{{\bf p},\,\omega}=\\ \dfr{\gamma}{\mu-i\gamma\omega}\sum\limits_{n=0}^{\infty}\left(\dfr{{\bf p}^2\gamma G^R_{{\bf p},\,\omega}}{\mu-i\gamma\omega}\right)^n
=\dfr{\gamma}{\mu-{\bf p}^2G^R_{{\bf p},\,\omega}-i\gamma\omega},
\end{multline*}
where $G^R$ is the retard part of the full ${\bf A}$-field Green functions matrix.
Therefore, the Keldysh part of the time-dependent full correlation function matrix is
\begin{gather}\label{DSF}
S(\omega)\propto \Delta^K_{{\bf p},\,\omega}=\dfr{2\gamma\beta^{-1}}{(\mu-{\bf p}^2G^R_{{\bf p},\,\omega})^2+\gamma^2\omega^2}.
\end{gather}
Substituting the renumbered correlation function of the field ${\bf A}$, $G^R_{{\bf p},\,\omega}$, into this expression (see Appendix VII) allows us to obtain the frequency dependence of the dynamic structure factor shown in Fig.\,\ref{BP}.
In the figure, one can see the obtained function qualitatively exactly reproduces the character of the experimental frequency dependencies of the dynamic structure factor.  The frequency peak corresponds to the intermolecular mode band. Its amplitude increases dramatically with increasing temperature and the maximum shifts to low frequencies as expected for a collision-induced band. As can be expected, the vortex--vortex mutual scattering,  being described by the  $\Sigma_{{\bf p},\,\omega}$ term corresponds to the collective or cooperative motion effect in which the intermolecular oscillation modes dominate. Exactly this contribution defines the Boson peak presence in the dynamic structure factor.

\section{Conclusions}

In conclusion one summarize the main thesis this work: the glass transition is a topological phase transition. The key condition for the possibility of such phase transition is the existence of non-trivial topologically protected structural excitations in the system that play the role of quasiparticles, and the glass transition represents the condensation in the subsystem of these quasiparticles.

The presented results of the theoretical analysis of the critical behavior of the proposed model allow us to assert that it gives an adequate description of most of the universal properties inherent in glass transitions in various systems.
It describes the topological phase transition between disordered states of elastic condensed matter, characterizing by the changing of the static shear modulus from $\mu_{st}=0$ in the hight temperture state to $\mu_{st}\neq 0$ in the low temperature.  With this the modulus scales with the glass transition temperature $\mu_{st}=\mu(T_g) \propto T_g$, that correspond to the experimentally observed properties of glass transition.

It si shown that the distinctive feature of this phase transition, as the Berezinskii--Kosterlitz--Thouless transition in the two-dimensional XY-model, is the absence of an order parameter. The long-range translational order does not occur during this transition, and the phase transition manifests itself in an infinite increase in the correlation radius and relaxation time as the system approaches the critical point. However, the derived in framework of the model and experimentally observed divergence of the nonlinear susceptibility and a sharp peak in the temperature dependence of the heat capacity (in the case of an increase in temperature) confirm our conclusion that the glass transition is a special type of thermodynamic phase transition.

Since the interaction between quasiparticles (disclinations) is long-range the properties of the systems consisting of them are largely determined by cooperative effects, which is especially strong on the kinetics of the system near the critical point. We can observe this both in the anomalous divergence of the relaxation time of the system according to VFT law and in the appearance of a boson peak in the low-frequency region of a dynamic structural factor. In turn, an abnormal relaxation slowdown leads to such experimentally observed effects as the dependence of the glass transition temperature on the cooling rate.

Thus, with the help of the proposed model, it is possible to describe well both the thermodynamic and kinetic properties of glass-forming systems. Considering the above, it can be argued that the model presented in the paper and its analysis prove the topological nature of glass transition.

\section{Acknowledgments}

I thank V.\,N. Ryzhov, V.\,M. Vinokur, and V.\,V. Brazhkin for stimulating discussions.

\newpage

\section*{Appendix I}%(averaging over grand canonical ensemble)

The systems effective Hamiltonian has the following form:
\begin{gather*}
\mathcal{H}= \dfr{\beta^{-2}}{2\mu}(\nabla\times {\bf A})^2-i\beta^{-1}{\bf \Omega} {\bf A}\sum\limits_{n=1}^{N}J\delta_{{\bf r=r}_n}^{(2)}.
\end{gather*}
In order to take account of all possible vortices configurations, we carry out the averaging over a grand canonical ensemble of the ``particles'' endowed with the two possible dimensionless charges: $J_n=\pm 1$.
Then the path integral is:
\begin{multline*}
\displaystyle W=\int \mathcal{D}A\left\{\exp\left[-\beta\int \mathrm{d}^3{\bf r}\,\dfr{\beta^{-2}}{2\mu}(\nabla\times {\bf A})^2\right] \times\right.\\\left.
\displaystyle \sum\limits_{N=1}^{\infty}\dfr {(e^{-\beta\mathcal{E}_c})^N}{N!}\prod\limits_{n=1}^N \int\mathrm{d}^3{\bf r}_n\sum\limits_{J_n=\pm 1}
\exp\left[ iJ_n{\bf \Omega A}({\bf r}_n)\right]\right\}.
\end{multline*}
After summation the systems effective Hamiltonian density assumes the form:
\begin{gather*}
\mathcal{H}= \dfr{\beta^{-2}}{2\mu}(\nabla\times {\bf A})^2-g\beta^{-1}\cos\left({\bf \Omega  A}\right),
\end{gather*}
where  $g=e^{-\beta\mathcal{E}_c}$ is the vortices density, and which is nothing but the Hamiltonian density of the sine-Gordon theory.

\section*{Appendix II}%(vortices correlation)

Let us consider the correlation of the ${\bf J_{p}}$ vortices with some moment ${\bf p}$, when
\begin{gather*}
{\bf J_p}=\int \mathrm{d}^d{\bf r}\, \delta^{(2)} ({\bf r}-{\bf r}_n){\bf J}_n e^{i{\bf pr}}\approx J_n e^{i{\bf pr}_n}.
\end{gather*}
In this case the pair correlation function of the vortices is
\begin{multline*}
\langle {\bf J_p}{\bf J_{-p}}\rangle %=\langle\langle {\bf J_p}{\bf J_{-p}}\rangle_J\rangle_A
=\left<{\bf J_p}{\bf J_{-p}}\exp\left[-i {\bf \Omega} \left({\bf J_{-p}}{\bf A_p}+{\bf J_p}{\bf A_{-p}}\right)\right]\right>_A=
\\\dfr{{\bf J_p}{\bf J_{-p}}}{ \int\mathcal{D}{\bf A}\exp\left[-\beta H\right]}\int\mathcal{D}{\bf A}\exp\left[-\beta H +i {\bf \Omega} \left({\bf J_{-p}}{\bf A_p}+{\bf J_p}{\bf A_{-p}}\right)\right]=\\
{\bf J_p}{\bf J_{-p}}\exp\left[-{\bf \Omega}^2{\bf J_p}\langle {\bf AA}\rangle_{\bf p}{\bf J_{-p}} \right]
\propto \exp\left[-\dfr{{\bf \Omega}^2E\beta}{\mu^{-1}{\bf p}^2+M^2 }\right].
\end{multline*}

\begin{figure}[h!]
   \centering
   \includegraphics[scale=0.4]{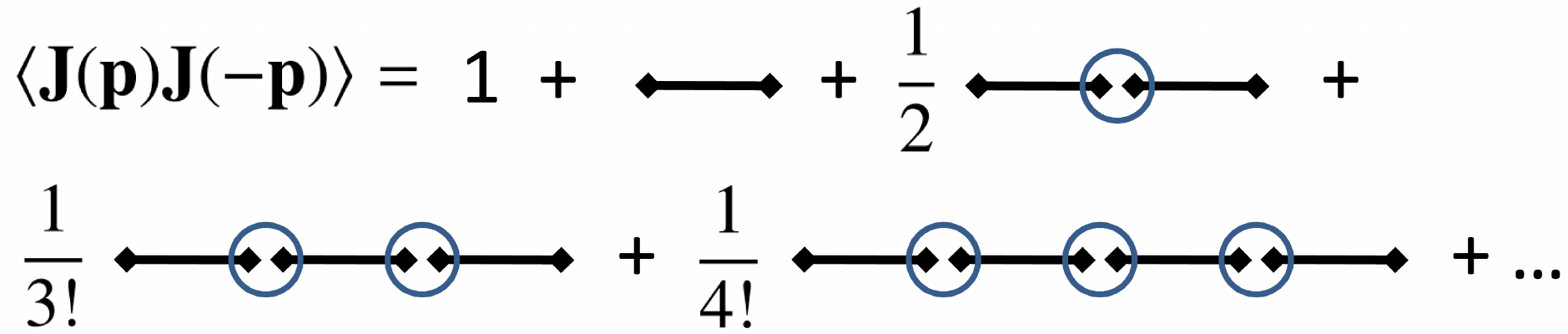}
   \caption{\textbf{Every dot corresponds to the $JA$ vortex, every circle corresponds to the free vertices correlation without interaction $\langle JJ\rangle_0=\delta({\bf r})$ ($\langle J({\bf p})J(-{\bf p})\rangle_0=1$).}}
  \label{cor}
\end{figure}

Another way to express the vortex correlator is the standard method by using an auxiliary field. In this case
the Hamiltonian (\ref{Z}) is rewritten  as follows:
\begin{gather*}
\mathcal{H}= \dfr{\beta^{-2}}{2\mu}(\nabla\times {\bf A})^2-i\beta^{-1}\left({\bf \Omega} {\bf A}+X\right) \sum\limits_{n=1}^{N}{\bf J}\delta_{{\bf r=r}_n}^{(2)},
\end{gather*}
and the correlator is expressed as
\begin{gather*}
\langle {\bf J J} \rangle_{\bf r-r'}=\left.-\dfr1W \dfr{\delta^2 W}{\delta X_{\bf r}\delta X_{\bf r'}}\right|_{X=0}.
\end{gather*}
After averaging over grand canonical ensemble of vortices the Hamiltonian is
\begin{gather*}
\mathcal{H}= \dfr{\beta^{-2}}{2\mu}(\nabla\times {\bf A})^2-g\beta^{-1}\cos\left({\bf \Omega A} +X \right).
\end{gather*}
Therefore
\begin{multline*}
\langle {\bf J J} \rangle_{\bf r-r'}=-g^2\langle\sin({\bf \Omega} {\bf A}_{\bf r})\sin({\bf \Omega} {\bf A}_{\bf r'})\rangle\\=-\dfr{g^2}2\langle\cos({\bf\Omega} ({\bf A}_{\bf r}-{\bf A}_{\bf r'}))-\cos({\bf \Omega} ({\bf A}_{\bf r}+{\bf A}_{\bf r'}))\rangle\\=
\dfr{g^2}2\exp\left[-\dfr{{\bf \Omega}^2}2 \langle ({\bf A}_{\bf r}+{\bf A}_{\bf r'})^2\rangle\right]-\dfr{g^2}2\exp\left[-\dfr{{\bf \Omega}^2}2 \langle ({\bf A}_{\bf r}-{\bf A}_{\bf r'})^2\rangle\right]\\=
g^2\exp\left[-\dfr{{\bf \Omega}^2}2\langle {\bf A}^2\rangle -\dfr{\Omega^2}2\langle {A}^2\rangle\right]\sinh\left[-{\bf \Omega}^2 \langle {\bf AA}\rangle_{\bf r-r'}\right]\\\approx -\dfr{g^2}2\exp\left[-{\bf \Omega}^2\langle {\bf A}^2\rangle\right]\exp\left[{\bf \Omega}^2\langle {\bf AA}\rangle_{\bf r-r'}\right].
\end{multline*}
In the pulse representation this correlator is written as follows:
\begin{multline*}
\langle {\bf J J} \rangle_{{\bf p}=0}=\int\mathrm{d}^3{\bf r}  \langle {\bf J J} \rangle_{\bf r}\propto \int\mathrm{d}^3{\bf r}\exp\left[{\bf \Omega}^2\langle {\bf A A}\rangle_{\bf r}\right]\\
=\int\mathrm{d}^3{\bf r}\left[1+{\bf \Omega}^2\langle {\bf A A}\rangle_{\bf r}+\dfr{{\bf \Omega}^4}2\langle {\bf A A}\rangle_{\bf r}^2+\ldots \right]\\
=1+{\bf \Omega}^2\langle {\bf A A} \rangle_{{\bf p}=0}+\dfr{{\bf \Omega}^4}2\int\dfr{\mathrm{d}^3{\bf p'}}{(2\pi)^3}\langle {\bf A A}\rangle_{\bf p'}^2+\ldots\\
\approx \exp\left[{\bf \Omega}^2\langle {\bf A A} \rangle_{{\bf p}=0}\right].
\end{multline*}

\section*{Appendix III}  %(correlation length)

If we suppose that $|\langle {\bf J}{\bf J}\rangle|_{\bf r}\propto \exp (-|{\bf r}|/L_c)$, then
\begin{multline*}
\int \mathrm{d}^3{\bf r}|\langle {\bf J}{\bf J}\rangle|_{\bf r}\\
\propto\int \mathrm{d}^3{\bf r} \exp (-|{\bf r}|/L_c)\\
=4\pi \int \mathrm{d}r r^2\exp (-r/L_c)=4\pi L_c^3\int \mathrm{d}x x^2\exp (-x)\propto L_c^3.
\end{multline*}
Then again,
\begin{gather*}
\int \mathrm{d}^3{\bf r}|\langle {\bf J}{\bf J}\rangle|_{\bf r}=\int \mathrm{d}^3{\bf r} \int\limits_0^{1}\dfr{\mathrm{d}^3{\bf p}}{(2\pi)^3}|\langle {\bf J}{\bf J}\rangle|_{\bf p}e^{i{\bf pr}}=\\
\int\limits_0^{1}\mathrm{d}^3{\bf p}|\langle {\bf J}{\bf J}\rangle|_{\bf p}\delta^{(3)}({\bf p})=|\langle {\bf J}{\bf J}\rangle|_{p=0}.
\end{gather*}
Therefore $L_c\propto|\langle{\bf J}{\bf J}\rangle|_{p=0}^{1/3}$.

\section*{Appendix IV} %(relationship between ${\bf u}$-field and ${\bf A}$-field correlators)

After functional integration of the model with the Hamiltonian (\ref{S1}) over $\hat{\bf u}$ field the systems effective Hamiltonian takes the following form:
\begin{gather*}
\mathcal{H}= \dfr{\beta^{-2}}{2\mu}(\nabla\times {\bf A}-i\beta\hat{\bf \sigma})^2-i\beta^{-1}{\bf \Omega} {\bf A}\sum\limits_{n=1}^{N}J\delta_{{\bf r=r}_n}^{(2)},
\end{gather*}
The sought-for correlation functions are found by the differentiation of the partition function over $\hat{\bf \sigma}$ field.
The pair correlation function is derived by the double differentiation:
\begin{multline*}
\langle{\bf u}{\bf u}\rangle_{\bf p}=-\dfr1{{\bf p}^2}\dfr {\beta^{-2}}{W}\left.\dfr{\delta^2 W}{\delta \hat{\bf \sigma}_{\bf p}\delta \hat{\bf \sigma}_{\bf -p}}\right|_{\hat{\bf \sigma}=0}\\
=\dfr{1}{\beta \mu}\left(\dfr{1}{\beta \mu}\langle {\bf AA}\rangle_{\bf p} +\dfr {1}{{\bf p}^2}\right),
\end{multline*}
and the quadratic correlation function is determined as:
\begin{multline*}
\langle{\bf u}^4\rangle_{\bf p}=-\dfr1{{\bf p}^4}\dfr {\beta^{-4}}{W}\left.\dfr{\delta^4 W}{\delta \hat{\bf \sigma}_{{\bf p}}\delta \hat{\bf \sigma}_{{\bf p}}\delta \hat{\bf \sigma}_{{\bf p}}\delta \hat{\bf \sigma}_{{\bf p}}}\right|_{\hat{\bf \sigma}=0}\\
=\left(\dfr{1}{\beta \mu}\right)^2\left(\left(\dfr{1}{\beta \mu}\right)^2\langle {\bf A}^4\rangle_{\bf p}+3\left(\dfr{1}{\beta \mu}\right)^2\langle {\bf AA}\rangle^2_{\bf p}+\right.\\\left.
6\dfr{1}{{\bf p}^2}\dfr{1}{\beta \mu}\langle {\bf AA}\rangle_{\bf p}+3\dfr{1}{{\bf p}^4}\right)\\
=\left(\dfr{1}{\beta\mu}\right)^4\langle {\bf A}^4\rangle_{{\bf p}=0}+
3\left(\dfr{1}{\beta\mu}\left[ \dfr{1}{\beta\mu}\langle {\bf AA}\rangle_{\bf p}+\dfr1{{\bf p}^2}\right]\right)^2 \\
=\left(\dfr{1}{\beta\mu}\right)^4\langle {\bf A}^4\rangle_{{\bf p}}+
3 \langle{\bf u}{\bf u}\rangle^2_{\bf p}.
\end{multline*}

\section*{Appendix V}   % (heat capacity)

According to the heat capacity definition the contribution to this value of the vortex subsystem is
\begin{multline*}
\Delta C_p=\dfr1{\beta}\dfr{\partial^2\left( T\ln W\right)}{\partial T^2}\\
=\dfr1{\beta}\left(\dfr2W\dfr{\partial W}{\partial T}-T\left(\dfr1W \dfr{\partial W}{\partial T}\right)^2+T\left(\dfr1W \dfr{\partial^2 W}{\partial T^2}\right)\right).
\end{multline*}
By substituting the partition function $W$ in this expression one can estimate the heat capacity near $T^*$. At the approach to the $T^*$ from below the  heat capacity is written as follows:
\begin{multline*}
\Delta C_p\approx \dfr{k_bT^2}W \dfr{\partial^2 W}{\partial T^2}=\dfr{T}2\dfr{\partial^2}{\partial T^2}\int\dfr{\mathrm{d}^3{\bf p}}{(2\pi)^3}\mu\langle{ \hat{\bf u}\hat{\bf u}}\rangle_{\bf p}\\ =
\dfr{T}2\dfr{\partial^2}{\partial T^2}\int\dfr{\mathrm{d}^3{\bf p}}{(2\pi)^3} \left(\dfr {{\bf p}^2(\beta\mu)^{-1}}{\mu^{-1}{\bf p}^2+\beta^2M ^2}+\beta^{-1}\right)\\ \approx \dfr{T}{2\beta\mu}\dfr{\partial^2}{\partial T^2}\int\dfr{\mathrm{d}^3{\bf p}}{(2\pi)^3}\dfr{-M^2}{\mu^{-1}{\bf p}^2+\beta^2M ^2}\propto \dfr1{|T^*-T|^{1/2}}.
\end{multline*}

\section*{Appendix VI}

In order to calculate the following integral:
\begin{multline*}
I^A(t)=\iint \dfr{\mathrm{d}^3{\bf p'}\mathrm{d}\omega'}{(2\pi)^4}\dfr{\mathrm{d}^3{\bf p''}\mathrm{d}\omega''}{(2\pi)^4}\times\\
G^K_{{\bf p'},\,\omega'}G^K_{{\bf p''},\,\omega''}G^{A}_{{\bf p'}+{\bf p''}+{\bf p},\,\omega'+\omega''+\omega},
\end{multline*}
it is convenient to consider it in the $({\bf k},\,t)$ presentation:
\begin{multline*}
I^A(t)=\beta^{-2}\iint \dfr{\mathrm{d}^3{\bf k}}{(2\pi)^3}\dfr{\mathrm{d}^3{\bf k'}}{(2\pi)^3}\theta\left(t(\mu^{-1}({\bf k}+{\bf k'})^2+M^2)\right)\times\\
\dfr{\exp\left[-2\gamma\mu^{-1}({\bf k}^{2}-{\bf k'}{\bf k}+{\bf k'}^2)t-3M^2t\gamma^{-1}\right]}{(\gamma^{2}\mu^{-1}{\bf k}^{2}+M^2)(\gamma^{2}\mu^{-1}{\bf k'}^{2}+M^2)}.
\end{multline*}
It can be integrated as follows:
\begin{multline*}
I^A(t)=\beta^{-2}\left(\dfr{\mu}{\gamma^2}\right)^3\dfr{(2\pi)^2}{(2\pi)^6}
\int\limits_{0}^1\mathrm{d}X\times\\
\iint\limits_{-\infty}^{\infty} \mathrm{d}z \mathrm{d}z'\theta\left(t(z^{2}-2z'zX+z'^2+M^2)\right)\times\\
\dfr{z^2z'^2\exp\left[-2\gamma^{-1}(z^{2}-z'zX+z'^2)t-3M^2t\gamma^{-1}\right]}{(z+iM)(z-iM)(z'+iM)(z'-iM)}=\\
\beta^{-2}\left(\dfr{\mu}{\gamma^2}\right)^3\dfr{(2\pi)^2}{(2\pi)^6}
\int\limits_{0}^1\mathrm{d}X\int\limits_{-\infty}^{\infty} \mathrm{d}z \,\theta\left(t(z^{2}-2iMzX)\right)\times\\
\dfr{-2\pi i M^2 z^2\exp\left[-2(z^{2}-iMzX-M^2)t\gamma^{-1}-3M^2t\gamma^{-1}\right]}{2iM(z+iM)(z-iM)}=\\
\beta^{-2}\left(\dfr{\mu}{\gamma^2}\right)^3\dfr{(2\pi)^2}{(2\pi)^6}\int\limits_{0}^1\mathrm{d}X\pi^2 M^2 \theta\left(t(2X-1)\right)\times\\
\exp\left[-2(M^2X-M^2)t\gamma^{-1}-M^2t\gamma^{-1}\right]=\\
\beta^{-2}\left(\dfr{\mu}{\gamma^2}\right)^3\dfr{\pi^2 M^2 }{(2\pi)^4}\int\limits_{0}^1\mathrm{d}X\,\theta\left(t(2X-1)\right)\times\\
\exp\left[-(2M^2X-M^2)t\gamma^{-1}\right]=\\
\beta^{-2}\left(\dfr{\mu}{\gamma^2}\right)^3\dfr{1}{(2\pi)^2}\dfr{\gamma}{8t}\left(\theta(-t)\exp\left[M^2t\gamma^{-1}\right]\times \right.\\\left.
-\theta(t)\exp\left[-M^2t\gamma^{-1}\right]\right).
\end{multline*}
We take into account that $M^2\ll 1$ and $\gamma >t$, thus
\begin{gather*}
I^A(t)\approx -\theta(t)\beta^{-2}\left(\dfr{\mu}{\gamma^2}\right)^3\dfr{\gamma}{8(2\pi)^2}\dfr{1}{t}.
\end{gather*}
After Fourier transformation
\begin{gather*}
I^A(\omega)\approx \beta^{-2}\left(\dfr{\mu}{\gamma^2}\right)^3\dfr{\gamma}{8(2\pi)^2}\left(\log\omega+\mathcal{C}\right),
\end{gather*}
where $\mathcal{C}$ is Euler's constant which can be neglected at small $\omega$.

\section*{Appendix VII}

Near the glass transition the ${\bf A}$-field full correlation function matrix, $\hat G_{{\bf p},\,\omega}$, differs from the prime one by the addition of the self-energy term,
\begin{gather*}
\Sigma_{{\bf p}\approx 0,\,\omega}\approx M^4\dfr{\mu^3}{\beta^2\gamma^7}\dfr{\Omega^4}{8(2\pi)^2}\log\omega,
\end{gather*}
to the ${\bf A}$-field mass: $M^2\,\to\,{M_R^2}_{{\bf p},\,\omega}=M^2-\gamma\Sigma_{{\bf p},\,\omega}$.
\begin{multline*}
G^R_{{\bf p},\,\omega}={G^A}^*_{{\bf p},\,\omega} =\dfr{\gamma}{\gamma^2\mu^{-1}{\bf p}^2+M^2-\gamma\Sigma_{{\bf p},\,\omega}-i\gamma \omega}\\
=\dfr{\gamma(\gamma^2\mu^{-1}{\bf p}^2+{M_R^2}_{{\bf p},\,\omega})}{(\gamma^2\mu^{-1}{\bf p}^2+{M_R^2}_{{\bf p},\,\omega})^2+\gamma^2 \omega^2}+\\
\dfr{i\gamma^2 \omega}{(\gamma^2\mu^{-1}{\bf p}^2+{M_R^2}_{{\bf p},\,\omega})^2+\gamma^2 \omega^2},\\
G^K_{{\bf p},\,\omega}=\dfr{2\gamma\beta^{-1}}{(\gamma^2\mu^{-1}{\bf p}^2+{M_R^2}_{{\bf p},\,\omega})^2+\gamma^2 \omega^2}.
\end{multline*}
As a result the ${\bf u}$-field full correlation functions have the following form:
\begin{multline*}
\Delta^R_{{\bf p},\,\omega}={\Delta^A}^*_{{\bf p},\,\omega}=\langle {\hat{\bf u}\hat{\bf u}'}\rangle_{{\bf p},\,\omega}= \dfr{\gamma}{\mu-i\gamma\omega}\sum\limits_{n=0}^{\infty}\left(\dfr{{\bf p}^2\gamma G^R_{{\bf p},\,\omega}}{\mu-i\gamma\omega}\right)^n
\\
=\dfr{\gamma}{\mu-{\bf p}^2\gamma G^R_{{\bf p},\,\omega}-i\gamma\omega},
\end{multline*}
\begin{multline*}
\Delta^K_{{\bf p},\,\omega}= \dfr{2k_bT}{\gamma \omega}\Im\left(\Delta^R_{{\bf p},\,\omega}\right)=\\
2\gamma k_bT\left[\omega^2\gamma^2\left(1+{\bf p}^2\dfr{\beta\gamma}2G^K_{{\bf p},\,\omega}\right)^2+  \right.\\ \left.
\left(\mu-{{\bf p}^2(\gamma^2\mu^{-1}{\bf p}^2+{M_R^2}_{{\bf p},\,\omega})\dfr{\beta\gamma}2G^K_{{\bf p},\,\omega}}\right)^2\right]^{-1}.
\end{multline*}

\end{document}